# Photon Spacecraft and Aerocapture: Enabling Small Low-Circular Orbiters at Mars and Venus

Athul Pradeepkumar Girija [1,**,**]

[1]*School of Aeronautics and Astronautics, Purdue University, West Lafayette, IN 47907, USA*

**ABSTRACT**

With advancements in low-cost launchers and small interplanetary spacecraft, NASA has recognized the potential of small missions to perform focused planetary science investigations at Mars and Venus. The EscaPADE, part of the NASA SIMPLEx program will deliver two small spacecraft to elliptical orbits around Mars using the Photon spacecraft. Orbit insertion, particularly to low-circular orbits requires significant propellant, taking up a substantial fraction of the Photon wet mass and present a significant challenge for small missions. The large ΔV requirements for low-circular orbit make it difficult to insert small satellites into these orbits even with the highly capable Photon, as the total ΔV for Earth escape and orbit insertion exceeds its capability. Drag modulation aerocapture offers a promising alternative, using the atmospheric drag to obtain the large ΔV. The study shows how the Photon when combined with drag modulation aerocapture can deliver small orbiters to low-circular orbits, enabling a wide range of small orbiter missions. Aerocapture eliminates the need for Photon to provide 2 to 3.5 km/s of ΔV for orbit insertion, which translate into mass and cost savings, and can enable frequent low-cost small orbiters and small satellite constellations at Mars and Venus in the near future.

***Keywords:*** Photon, Aerocapture, Low-Cost Mission, Mars, Venus

[****] To whom correspondence should be addressed, E-mail: athulpg007@gmail.com

## I. INTRODUCTION

With advancements in low-cost launchers and small interplanetary spacecraft, NASA has recognized the potential of low-cost missions to perform focused planetary science investigations at Mars and Venus in the near future [1]. The Escape and Plasma Acceleration and Dynamics Explorers (EscaPADE), part of Small Innovative Missions for Planetary Exploration (SIMPLEx) program mission will deliver two identical 200 kg spacecraft to Mars in 2024 [2]. The Photon is first delivered to low-Earth orbit, where it uses its engines to perform a series of orbit raising burns, and then performs a trans-Mars injection burn. Upon arrival at Mars, the Photon burns again to capture into orbit and then performs orbit reduction maneuvers to deliver the two spacecraft into their 160 x 8400 km science orbit. At a cost of only $79 million including the launch vehicle, EscaPADE will be the first demonstration of standalone low-cost interplanetary orbiter mission. Rocket Lab has also made the commitment to fly a private mission to deliver a 20 kg probe for in-situ sampling of the Venusian clouds, also using the Photon high-performance spacecraft (shown in Figure 1) with an estimated budget under $10 million [3]. Orbit insertion, particularly to low-circular orbits requires significant propellant and can require about 20–40% of the Photon wet mass and present a significant challenge for small missions [4, 5]. Aerocapture can be used to eliminate the substantial propellant need for orbit insertion [6, 7]. The present study shows how the Photon spacecraft when combined with drag modulation aerocapture, can enable a wide range of low-cost small orbiter missions and small satellite constellations at Mars and Venus in the near future.

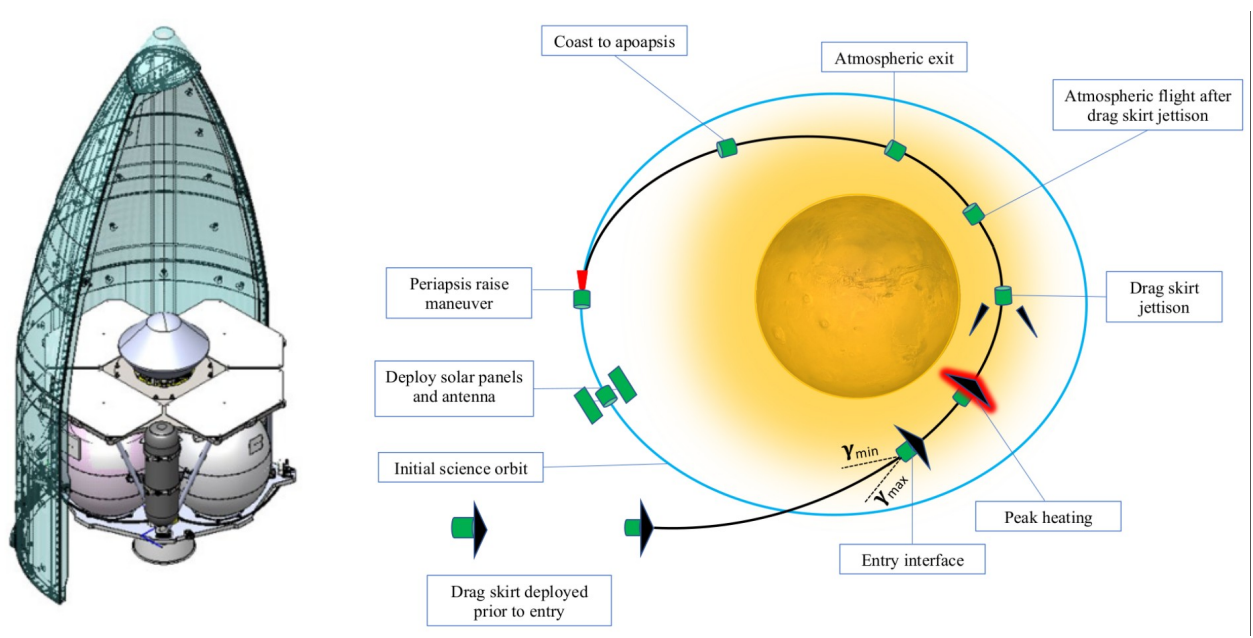

Figure 1. (Left) Photon upper stage with the Venus probe inside the Electron rocket fairing, courtesy of Rocket Lab. (Right) Schematic of the drag modulation aerocapture concept.

## II. CHALLENGE OF ORBIT INSERTION

The high-energy Photon is a highly-capable spacecraft which can provide nearly 3 km/s of ΔV. The EscaPADE mission for example uses about 1.1 km//s for the orbit raise and escape burns combined, and about 1.3 km/s for orbit insertion and reduction to a 165 x 6400 km elliptical orbit. For the EscaPADE mission, the Photon spacecraft weighs approximately 500 kg wet when delivered to LEO, of which about 150 kg is used for orbit raise and Earth escape. Mars orbit insertion and reduction consumes an additional 130 kg of propellant, which leaves about 200 kg in the elliptical science orbit. However, orbit insertion to low-circular orbits require significantly more ΔV, approaching 2 km/s at Mars and 3.5 km/s at Venus as seen in Figure 2. The large ΔV requirements for low-circular orbit make it difficult to insert small satellites into these orbits even with the highly capable Photon spacecraft as the total ΔV for Earth escape and orbit insertion exceeds its capability. Drag modulation aerocapture offers a promising alternative, using the atmospheric drag to obtain the large ΔV with almost no propellant [8, 9]. Instead of the Photon spacecraft capturing itself into orbit with propulsion, it will release one or more small drag modulation aerocapture vehicles once it approaches the planet's sphere of influence. The small spacecraft then independently enters the Martian atmosphere and performs aerocapture. Thus the Photon spacecraft only needs a propulsion system big enough to perform the Earth escape and correction maneuvers (about 1.2 km/s), while the 2–3.5 km/s required for orbit insertion is obtained from aerocapture. The mass savings from not using propulsion for orbit insertion can be used to accommodate more science payload, or alternatively used to realize a smaller and cheaper Photon with significantly less propellant.

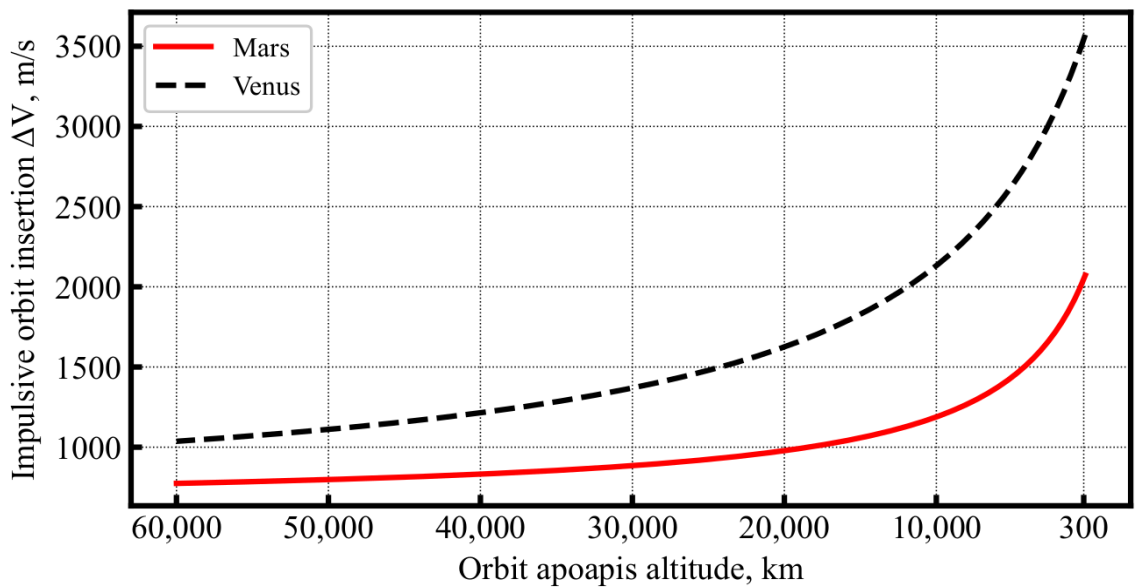

Figure 2. Orbit insertion ΔV at Mars and Venus, as function of the orbit apoapsis.

## III. AEROCAPTURE AT MARS

Insertion of a small satellite into a 200 x 400 km low-circular orbit at Mars is considered from the interplanetary trajectory used by the Mars 2020 mission. The orbit insertion ΔV is 2054 m/s. However, instead of performing a propulsive burn, the Photon targets the aim point for a aerocapture and releases the small satellite on a course for an entry trajectory within the drag modulation aerocapture corridor [-9.93, -8.96 deg]. Figure 3 shows the drag modulation aerocapture trajectory for entry at -9.1 deg, near the shallow limit [10]. During the aerocapture manuever, ΔV of 2086 m/s is obtained from atmospheric drag. At the first apoapsis, a 41 m/s periapsis raise burn is performed by the small satellite to achieve its desired 200 x 400 km orbit. The implications of the ΔV offered by aerocapture are discussed below. Without aerocapture, assuming 1.2 km/s for orbit raising and escape, and 2 km/s for orbit insertion, the total ΔV required to be supplied by the Photon is about 3.2 km/s. Though this is technically viable with the Photon, it puts it at the upper limit of the specifications even when not considering margins. With aerocapture, the Photon only needs to supply the 1.2 km/s required for Earth escape plus any margins, implying the Photon propulsion system can be much smaller and less expensive. When consdering the mass of the aerocapture system, the performance benefit compared to propulsion is only about 20%, but neverthless it may still be provide significant cost savings [11].

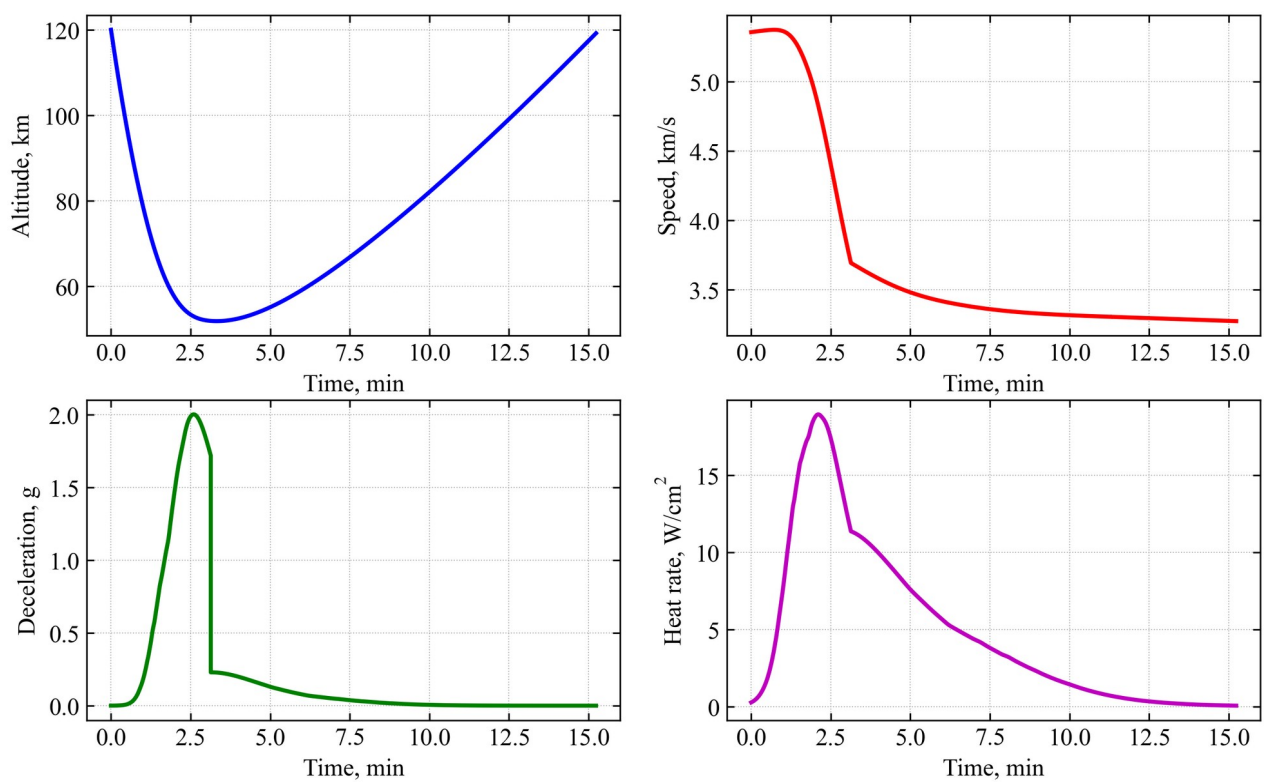

Figure 3. Drag modulation aerocapture trajectory at Mars.

## IV. AEROCAPTURE AT VENUS

Insertion of a small satellite into a 200 x 400 km low-circular orbit at Venus is considered from the interplanetary trajectory used by the Akatsuki mission. The orbit insertion ΔV is 3515 m/s, which is higher that what the Photon system can reasonably allocate for orbit insertion. Hence insertion into low-circular orbits at Venus is not viable with Photon. However, aerocapture can provide such large ΔV and thus can enable small satellite missions to Venus which require such low-circular orbits. The Photon will target the aerocapture entry corridor at [-5.54, -5.12] deg, and release the aerocapture vehicle on its approach trajectory and will then flyby Venus. Figure 3 shows the drag modulation aerocapture trajectory for entry at -5.2 deg, providing a ΔV of 3466 m/s. At the first apoapsis, a 30 m/s periapsis raise burn is performed by the small satellite to achieve its desired 200 x 400 km orbit. Without aerocapture, the total ΔV required to be supplied by the Photon is about 4.7 km/s. A recent NASA study assessing the cost drivers of small satellite missions found that ΔV is the largest driver of mission cost [12]. With aerocapture providing the 3.5 km/s, the Photon only needs to provide 1.2 km/s which translates into a much smaller Photon propulsion system design and significant cost savings. Thus the combination of Photon spacecraft which can launch independently on Electron, or as a secondary payload to GTO and aerocapture can enable frequent small low-cost Venus orbiter missions.

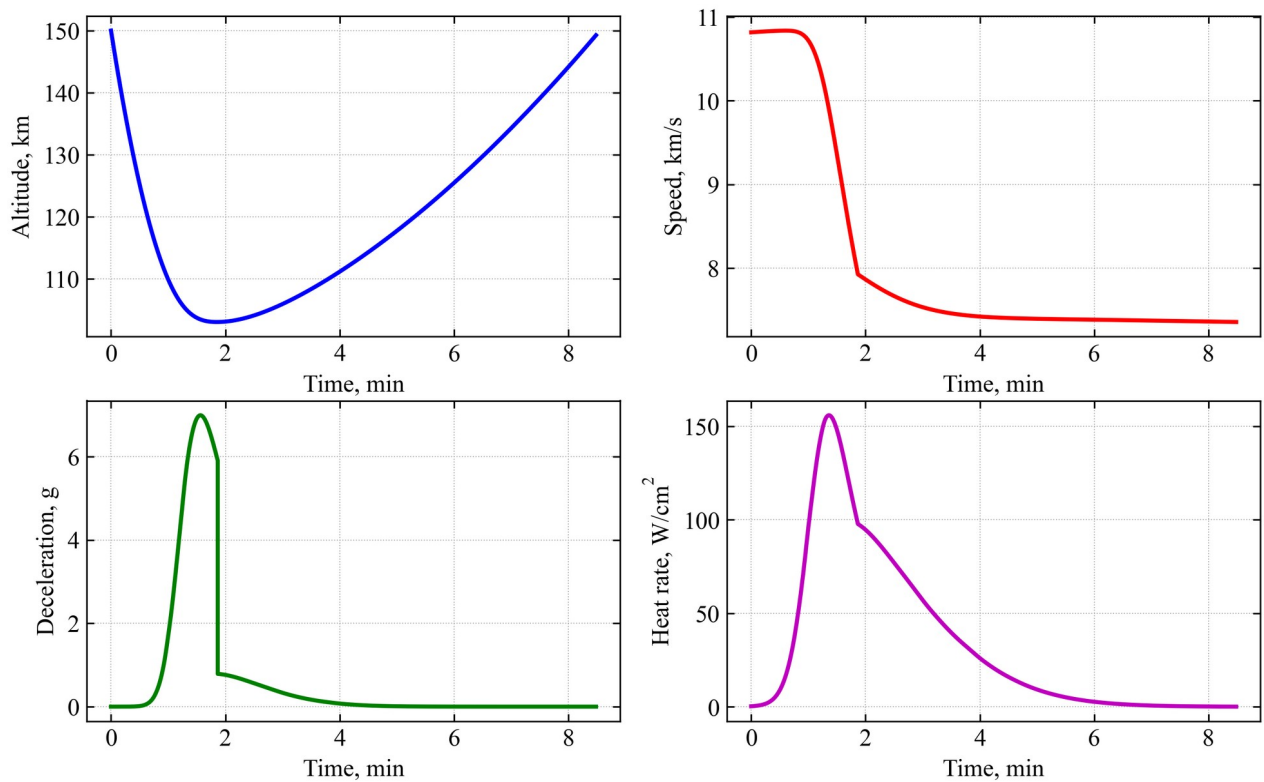

Figure 4. Drag modulation aerocapture trajectory at Venus.

## V. APPLICATIONS FOR FUTURE MISSIONS

The Photon spacecraft has demonstrated its viability for enabling small interplanetary science missions through the planned ESCAPADE and the Venus probe missions. When combined with drag modulation aerocapture, Photon offers even more capabilities for small low-cost orbiter missions, some of which are discussed in this section.

The ESCAPADE mission will demonstrate it is possible for small missions to enter Martian orbit using the Photon propulsion system. At Mars, the performance benefit in terms of delivered mass offered by aerocapture is small. However, it still has applications for mission such as an aerocapture technology demonstration. Due to the different atmospheric structure, Mars offers a much more benign aerothermal environment that Earth or Venus [13, 14]. The Photon can launch on an Electron or as a secondary payload to GTO. The Photon only needs about 1.2 km/s for Earth escape, and will act as a cruise stage carrying a small low-cost aerocapture technology demonstrator [15].

A single Photon can only deliver a single spacecraft to an orbit around a planet if its propulsion system is needed for orbit insertion. However, Photon when acting as cruise stage can carry multiple small satellites to form constellations around Mars and Venus. As seen in LEO, small imaging and radar satellite constellations can perform investigations previously done with large satellites at a fraction of the cost. By not having to do the orbit insertion burn, the Photon itself just needs enough propellant for the Earth escape and trajectory correction maneuvers. On approach the Photon can perform small divert manuevers to target each small satellite into their approach trajectories different inclination orbits. The different satellites will aerocapture into their respective orbits, while the Photon will flyby the planet. Considering an ESCAPADE like Photon which weighs 500 kg in LEO, and assuming about 150 kg of propellant is used for Earth escape, about 350 kg remains. Of this, assuming 100 kg is required for the Photon bus, this leaves about 250 kg for the small satellites. Assuming each small satellite weighs 50 kg with the aerocapture system mass included, this allows five satellites to be carried by a single Photon. Each of these satellites can be delivered to a different orbital plane or inclination using small divert manuevers by the Photon, enabling a constellation of five satellites to be established in a single launch. At Mars, these could be for example imaging satellites which can study the Martian surface or its climate. At Venus, these could be small radar satellites which can map the enitre surface.

At Venus, the combination of Photon and aerocapture is also applicable to larger missions which have the Photon as a secondary payload cruise stage for independent small satellites [16], and for low-circular orbits required for atmospheric sample return missions [17]. At Mars, in addition to a technology demonstration, it could enable frequent low-cost orbiter missions at a much faster cadence than the current missions. A technology demonstration mission at Mars will bring aerocapture into the realm of flight heritage after several decades of demonstration attempts [18].

In addition to near term missions at Mars and Venus, Rocket Lab also envisages small missions to the outer Solar System with the high-energy Photon [19]. Though the large distances and the need for radioisotope power make such small missions less realistic, it may become viable with advances in telecom and power system technologies. However, the demonstration of aerocapture at Mars or Venus will enhance its readiness for outer planet missions where its performance benefit is significantly greater [20, 21]. Even though aerocapture is not currently considered for the Uranus Orbiter and Probe which is the top priority for the Flagship mission of the next decade [22, 23], studies have shown aerocapture offers significant mission design advantages [24]. Aerocapture has also been shown to provide signficant benefits for a New Frontiers class Titan orbiter [25, 26], and a Neptune Flagship mission [27, 28].

## VI. CONCLUSIONS

The study showed how the Photon spacecraft when combined with drag modulation aerocapture can deliver small orbiters to low-circular orbits, enabling a wide range of small orbiter missions at Mars and Venus. Aerocapture eliminates the need for Photon to provide 2–3.5 km/s of ΔV for orbit insertion, which translates into mass and cost savings, and can enable small satellite imaging and radar constellations at Mars and Venus in the near future.

## DATA AVAILABILITY

The results presented in the paper can be reproduced using the open-source Aerocapture Mission Analysis Tool (AMAT) v2.2.22. The data and code used to make the study results will be made available by the author upon request.